\documentstyle[12pt,epsfig]{article}
\let\section=\subsection     \let\subsection=\subsubsection                

\begin{document}

\bigskip
\begin{center}
   {\large \bf THERMODYNAMICAL CHARACTERIZATION}\\[2mm]
   {\large \bf OF HEAVY ION REACTIONS}\\[5mm]
   T.~GAITANOS${}^{a}$, C.~FUCHS${}^{b}$, H.~H.~WOLTER${}^{a}$ \\[5mm]
   {\small \it ${}^{a}$Sektion Physik, Universit\"at M\"unchen \\
    D-85748 Garching, Germany \\[8mm] }
   {\small \it ${}^{b}$Institut f\"ur Theoretische Physik, Universit\"at T\"ubingen \\
    D-72076 T\"ubingen, Germany \\[8mm] }
\end{center}

\begin{abstract}\noindent
An detailed study of the thermodynamical state of nuclear matter in 
transport calculations of heavy--ion reactions is presented. 
In particular we determine temperatures from an analysis of the local 
momentum space distribution on one hand, and from a fit to fragment energy 
spectra in terms of a blast model with radial flow and temperature on the 
other. We apply this to spectator and participant matter. In spectator we 
find regions of spinodal instability with temperatures and densities which are 
consistent with experiments. In the participant we find different 
temperatures for different fragment masses, indicating that the fragments 
are not emitted from a source in thermal and chemical equilibrium.
\end{abstract}
\section{Introduction}
One of the challenges in the study of heavy--ion collisions is the 
understanding of multifragmentation in relation to liquid--gas 
phase transitions. In spite of many experimental and theoretical 
efforts these processes have not been fully understood yet. 
This is largely due to the fact that a heavy ion collision is 
a dynamical 
process where the state of nuclear matter varies strongly in space and time, 
and which during much of the 
reaction is not in global or even local equilibrium. 
This is, e.g. seen by looking at local momentum space distributions in 
transport calculations \cite{gait,erice} which are found to be highly 
anisotropic even during the compression phase of the collision. 
Only at the 
later stages of the reaction the local momentum distributions become 
more and more thermalized without neccessarily leading to global 
thermal equilibrium. Therefore, non--equilibrium effects are important for 
a reliable description of heavy ion collisions \cite{gait,essler}. 
The influence of these non--equilibrium effects on the determination of the 
equation of state of nuclear matter has been discussed in Ref. \cite{gait}. 
In this contribution we concentrate on the question of the applicability 
of thermodynamical concepts in the non--equilibrium situation of 
heavy ion collisions \cite{essler} and, 
in particular, with respect to phase transitions and multifragmentation. We 
want discuss the question whether, starting from a transport description of 
a heavy ion collision, where in principle everything is known about the 
system, a thermodynamical picture of multifragmentation can be deduced.
\section{Determination of temperature}
In this work we make use of two different methods of determining temperature: 
In the first, we determine local temperatures by fitting the local momentum 
distributions (obtained in our case from relativistic transport 
calculations)  to covariant 
Fermi--Dirac distributions at finite temperature in the local 
rest frame \cite{essler}. 
Non--equilibrium effects are taken into account by allowing a parametrization  
of the momentum space distribution in terms of two thermalized Fermi spheres 
\cite{gait}  (or covariantly  by ellipsoids). With this method we obtain a 
local microscopic temperature, $T_{loc}$. 

In the second method we follow  the experimental
method of fitting fragment energy spectra. These are generated in our 
calculations by applying a phase space coalescence algorithm 
\cite{gait} to the final stages of the transport calculations. 
As in experimental analyses these spectra are interpreted in a 
Siemens--Rasmussen \cite{sira} or blast model \cite{fopi}, which assumes a 
thermalized freeze--out configuration 
of nucleons and fragments with a collective radial flow profile and a 
unique temperature \cite{fopi,eos}. In this model the fragment spectra 
are given by 
\begin{eqnarray}
\frac{dN}{dE} & \sim & pE \int \: \exp(-\gamma E/T) \: 
\Bigg[ \frac{ \sinh{\alpha} }{ \alpha }\left( \gamma+\frac{T}{E} \right) - 
\frac{T}{E} \cosh{\alpha} \Bigg] \nonumber\\
& \times& n(\beta)\beta^{2} d\beta
\label{fit}
\qquad .
\end{eqnarray}
where $n(\beta)$ is the flow profile, $\gamma=\sqrt{1-\beta^{2}}$ and 
where T is the global temperature ($\alpha \equiv \gamma \beta p/T$). 
The flow profile is obtained from the simulation. Then the remaining 
parameter is the temperature $T$ which is fitted to experimental, 
resp. generated fragment spectra. In experimental analyses  
a global temperature is assumed, which characterizes the shapes of all fragment 
spectra. This is not obvious and should be clarified in the analyses of our 
transport calculations.
\section{Analysis of spectator matter (semi--central collisions)}

We first discuss the thermodynamical properties of spectator matter in 
semi--central Au on Au reactions at intermediate energies. This reactions 
has been studied extensively by the ALADIN collaboration \cite{aladin}. 
We determine the spectator temperature from fits to 
local momentum distributions (for more details see 
Ref. \cite{essler}). Fig. 1 shows the time evolution of the temperature 
in the spectator (left side) for different beam energies. When the 
spectators are clearly 
developed in the transport calculations after about 
$40$ fm/c, their temperatures  approach a rather constant value of 
about $T \approx 5$ MeV which remains fairly stable up to about 
80 fm/c, and furthermore is rather independent on the incident energy 
considered. 
\begin{center}
\begin{minipage}{13cm}
\baselineskip=12pt
\unitlength=1cm
\begin{picture}(12,5)
\put(0.0,-0.5){\makebox{\epsfig{file=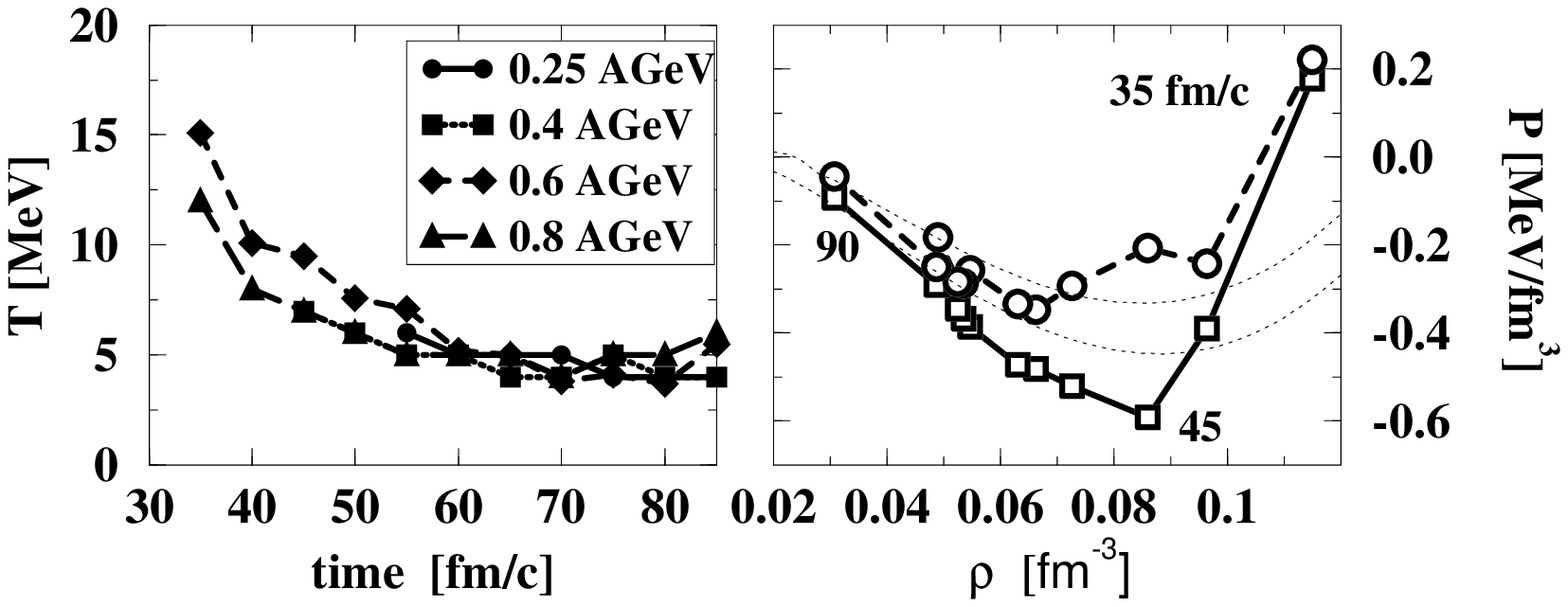,width=12.0cm}}}
\end{picture}
\\
{\begin{small}
Fig. 1. Left: Temperature evolution in the spectator in semi--central Au on Au 
reactions at different beam energies indicated in the figure. 
Right: Density-pressure trajectories for the spectator matter for the 
same reaction at 600 AMeV. The solid and dashed curves represent londitudinal and transverse pressure, respectively. The squares and circles are the values at different times 
starting from $t=35$ fm/c in steps of $5$ fm/c.  
The dotted curves are the nuclear matter isothermal equation of state 
for $T=5$ and $7$ (lower and upper curve, respectively).
\end{small}}
\end{minipage}
\end{center}
These results are in good agreement with experiments of 
the ALADIN collaboration, which from measurements 
with different "thermometers" determine the same value of $T \approx 5$ MeV,  
depending only moderately on the beam energy of the reaction. 
We also generate pressure--density trajectories for the spectator matter 
as a function of time 
(right side of Fig. 1). Dynamical instabilities should arise when  
the pressure increases with decreasing density indicating a 
negative effective compressibility, which occurs here at 
$t \geq 50$ fm/c. The system at this stage 
therefore enters an instability region and should break up into fragments. 
Comparing to the nuclear matter isothermal equation of state for temperatures of 
$T=5$ and $7$ MeV corresponding to the range of spectator temperatures 
in Fig. 1 one sees that the thermodynamical conditions, as determined here, 
are close to but not identical to those of equilibrated nuclear matter. Only at 
the final stages the spectator closely follows the nuclear matter behavior at 
temperatures of about $T \approx 5$ MeV. The densities at the instability 
condition are about $1/3-1/2$ of saturation density. It thus appears that the 
spectator closely approaches a freeze--out configuration in thermal and 
chemical equilibrium.

\section{Analysis of the fireball (central collisions)}
\begin{center}
\begin{minipage}{13cm}
\baselineskip=12pt
\unitlength=1cm
\begin{picture}(12,5)
\put(-0.3,-0.5){\makebox{\epsfig{file=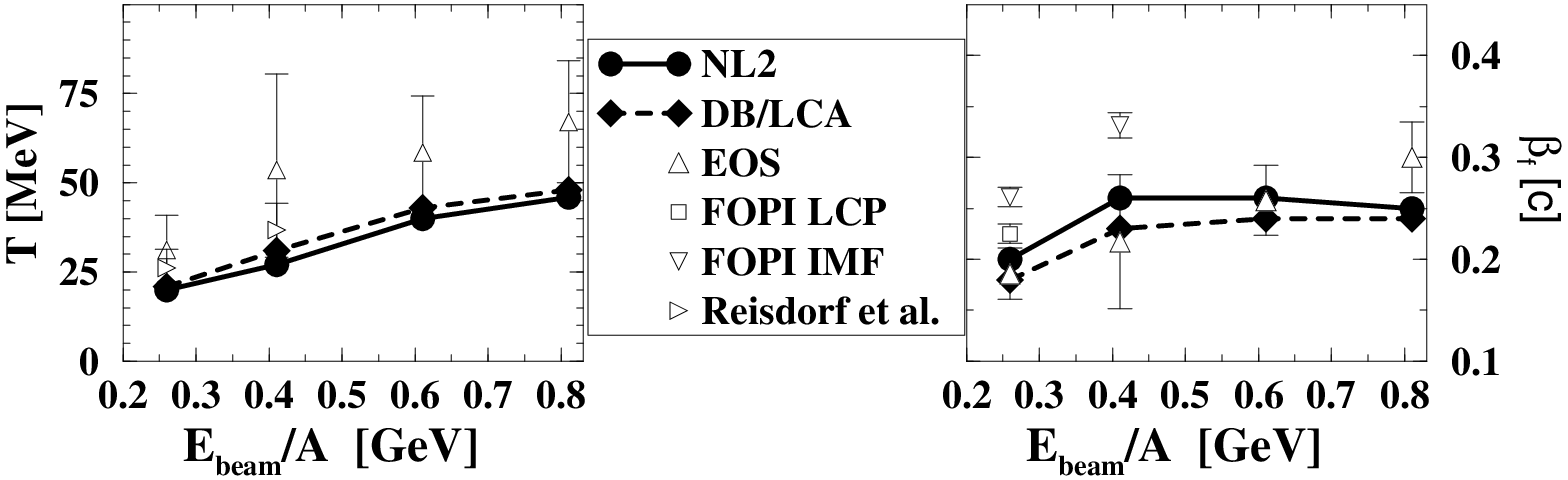,width=12.0cm}}}
\end{picture}
\\
{\begin{small}
Fig. 2. Temperatures (left) and radial flow (right) obtained from blast model 
fits to fragment energy spectra as function of the beam energy. 
The theoretical results are shown for two mean field models (see text). 
The data are taken from \protect\cite{fopi,eos}. 
\end{small}}
\end{minipage}
\end{center}
In central collisions the situation is rather different. If very central events 
are selected experimentally using 
charged particle multiplicities \cite{fopi} or theoretically  
at polar angles near mid rapidity \cite{eos}, there is no 
spectator matter. Rather one observes a hot dense fireball which expands isotropically as found by our calculations and also by other 
groups \cite{fopi}. Thus, 
assuming thermalization one can use Eq. (\ref{fit}) to extract the mean collective 
radial flow $\beta_{f}=<\beta>$ and a slope temperature $T_{slope}$ from fits to the fragment 
energy 
spectra. Fig. 2 shows the energy dependence of these quantities as 
determined from 
our calculations and from experiments for central 
Au on Au collisions. Two parametrizations of the mean field 
(non--linear Walecka model and configuration dependent Dirac--Brueckner 
mean fields \cite{gait}) were used in the calculations to demonstrate 
the moderate dependence of of 
$\beta_{f}$ and $T$ on the mean field. As seen in Fig. 2 the experimental 
data for the radial flow are reproduced very well. The 
comparison of the extracted slope temperatures $T_{slope}$ (left side of Fig. 2) is, 
however, only qualitative.

It is of interest, to discuss the relation of these slope temperatures to the 
local temperatures $T_{loc}$ determined from the momentum distribution of the 
calculation (we used a Maxwell--Boltzmann distribution here, in order 
to be consistent with eq. (\ref{fit}), but the difference is $\leq 5$ MeV in the 
final stages of the collision). It is also of interest to make the comparison 
separately for different fragment masses in order to determine whether a 
freeze--out szenario with a unique temperature is realistic. 
\begin{center}
\begin{minipage}{13cm}
\baselineskip=12pt
\unitlength=1cm
\begin{picture}(12,6)
\put(0.5,-0.5){\makebox{\epsfig{file=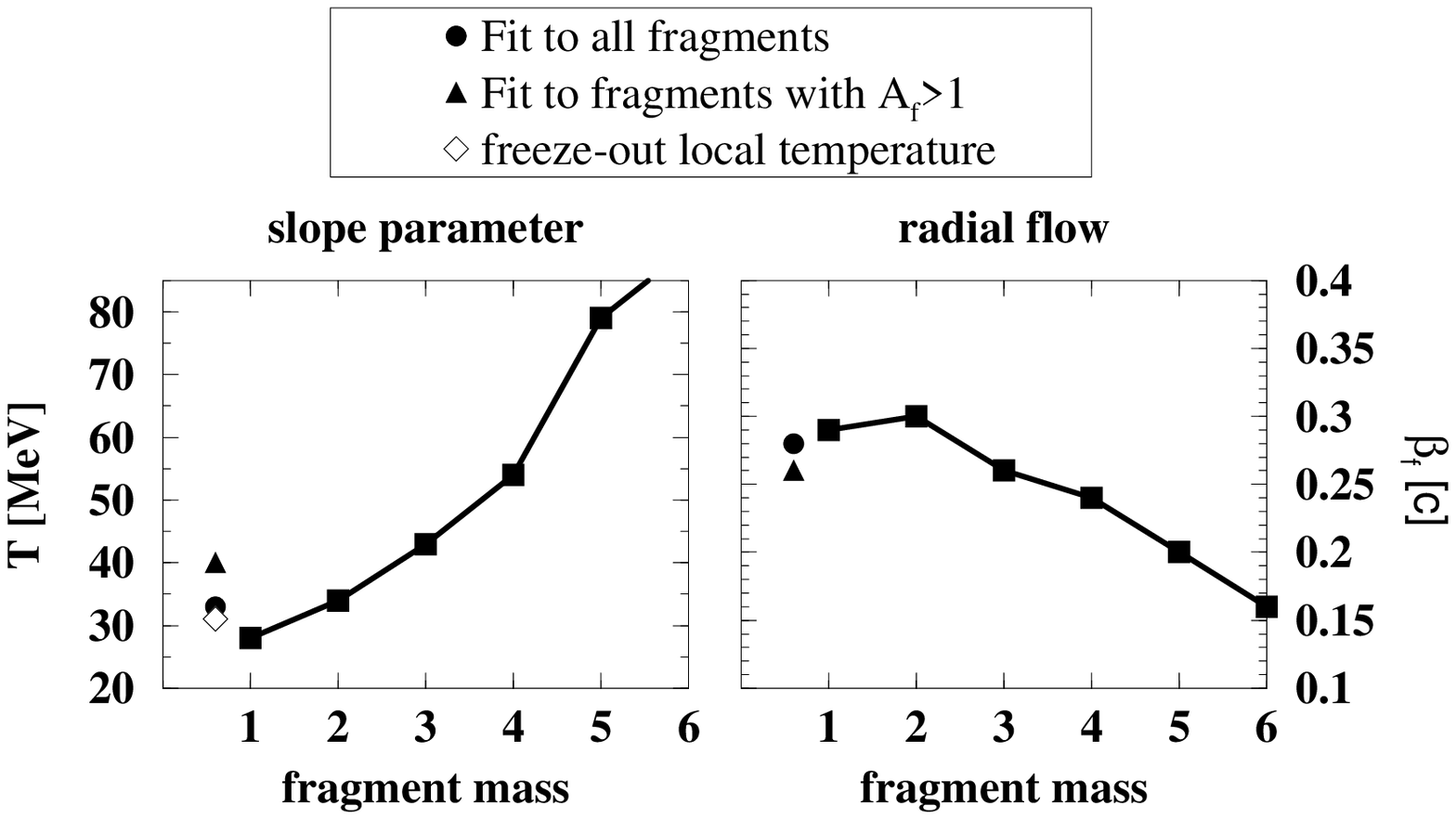,width=12.0cm}}}
\end{picture}
\\
{\begin{small}
Fig. 3. Temperatures (left) and radial flow (right) for fragments of different mass as determined from blast model fits. 
Also simultaneos fits to all fragments including nucleons (filled circle) 
and only to fragments with $A_{f} \geq 2$ (filled triangle) are shown. 
The open diamond in the left figure is the local temperature at freeze--out 
obtained from a fit to local momentum space distributions (see text).
\end{small}}
\end{minipage}
\end{center}
The results of this analysis are shown in Fig. 3. Here we show the slope 
temperatures and radial flow velocities from blast model fits to the spectra 
of different fragment masses $A_{f}$. The temperature increases and the radial flow 
decreases with increasing mass. In the coalescence picture such a behavior is 
reasonable, since a larger fragment has to be generated more inside the fireball, where 
the flow velocity is smaller and the temperature higher. In Fig. 3 we also 
show the result of a simultaneous blast model fit to all fragments and to fragments 
with mass $A_{f} \geq 2$. Since the fragment multiplicities are roughly 
exponential and thus dominated by the nucleons the results for all fragments 
are close to those for $A_{f}=1$ alone. On the other hand the fit to the heavier 
fragments alone has lower radial flow and higher temperature and is the one compared 
in Fig. 2 to the corresponding experimental value. 

Also shown in Fig. 3 is the local temperature from the momentum distributions 
determined at about $35$ fm/c. At this time the fireball in the 
calculations approaches a freeze--out configuration (nucleon--nucleon 
collisions cease) in equilibrium (pressure isotropic). It is then a consistent 
check that the local temperature for this situation agrees approximately 
with the one determined from a blast model fit to the $A_{f}=1$ energy spectra.

\section{Conclusions}
We have studied the thermodynamical state of nuclear matter in 
heavy ion collisions by 
analyzing local phase space configurations and by analyzing fragment 
energy spectra. For the spectator a consistency with temperatures and 
breakup conditions with results of the ALADIN 
collaboration was found. For the participant matter we have applied in 
addition a blast model analysis to fragment spectra generated in the 
coalescence model. We see that the slope temperatures in such a description 
do not yield a unique value for all fragments. This does not favour the 
picture of a freeze--out configuration in thermodynamical equilibrium. Rather 
it appears that fragment emission is a dynamical process which occurs during 
a longer stage of the heavy ion reaction. 


\end{document}